\documentclass[conference]{IEEEtran}
\IEEEoverridecommandlockouts

\usepackage{cite}
\usepackage{graphicx}
\usepackage{psfrag} 
\usepackage{bm}
\usepackage{subfigure}
\usepackage{amsmath,amssymb,amsfonts}
\usepackage{epsfig}
\usepackage{amsmath}
\usepackage{multirow}
\usepackage{lineno}
\usepackage{algorithm}
\usepackage{algorithmic} 
\usepackage{diagbox} 
\usepackage[margin=0.7in]{geometry}

\title{Reconfigurable AI Modules Aided Channel Estimation and MIMO Detection}

\author{Xiangzhao Qin$^{1}$, Sha Hu$^1$, Jiankun Zhang$^2$, Jing Qian$^2$, and Hao Wang$^2$ \\
 $^1$ Lund Research Center, Huawei Technologies Sweden AB, Sweden.\\ 
 $^2$ Huawei Technologies Co., Ltd., Beijing, China.  \\    
 Email: \{chrisqin, hu.sha, zhangjiankun4, qianjing3, hunter.wanghao\}@huawei.com
} 

\begin{document}
\maketitle

\begin{abstract}
Deep learning (DL) based channel estimation (CE) and multiple input and multiple output detection (MIMODet), as 
two separate research topics, have provided convinced evidence to demonstrate the effectiveness and robustness 
of artificial intelligence (AI) for receiver design. However, problem remains on how to unify the CE and MIMODet 
by optimizing AI's structure to achieve near optimal detection performance such as widely considered QR with 
M-algorithm (QRM) that can perform close to the maximum likelihood (ML) detector. In this paper, we propose an 
AI receiver that connects CE and MIMODet as an unified architecture. As a merit, CE and MIMODet only adopt 
structural input features and conventional neural networks (NN) to perform end-to-end (E2E) training offline. 
Numerical results show that, by adopting a simple super-resolution based convolutional neural network (SRCNN) 
as channel estimator and domain knowledge enhanced graphical neural network (GNN) as detector, the proposed 
QRM enhanced GNN receiver (QRMNet) achieves comparable block error rate (BLER) performance to 
near-optimal baseline detectors.   
\end{abstract}

\section{Introduction}
Channel estimation (CE) and MIMO detection (MIMODet), as the essential cores of communication problems, 
have attracted a lot of attention and their revolutions span from 2G to 5G and beyond. As the new 
requirements of 6G, especially for the user equipment (UE) communications with limited antenna size, 
challenges exist for conventional CE and MIMODet to support massive machine-type communication and 
ultra-reliable low latency communication (URLLC) with regards to performance improvement and complexity 
reduction. Recently, artificial intelligence (AI), specifically neural network (NN) enhanced deep learning (DL) 
approach, has been seen as a potential enabler for the next generation wireless system. Within the scope of 
CE and MIMODet, emerging solutions incorporate naive AI modules to replace the functionality of classical 
communication components such as CE \cite{Mehran2019Deep,Jiang2021Dual} and MIMODet\cite{Sun2020Learn} separately.  

In a data-driven approach, AI's output layer predicts the channel's coefficients 
or the soft-decisions of payloads in the forward propagation phase, and the weights of neurons are updated 
via back propagation. However, in the presence of dynamic environment over one OFDM subframe, e.g., within the 
scope of mm-Wave and Terahertz channels, it becomes extremely difficult to analytically model the underlying 
behaviour by using a shallow AI structure to perform E2E learning\cite{Zhou2021RCNet}. By enlarging 
the size of neurons, or designing the AI's structure deliberately, limited gain can be obtained, 
but its overall performance only competes with simple baselines \cite{Honkala2021DeepRx, Faycal2022E2E}, 
which is far from that of near-optimal solution such as QRM detection. To overcome the 
``uncertainty'' brought from neurons, recent works demonstrate the effectiveness of utilizing model driven 
DL approach to optimize the parameters of belief propagation (BP) algorithm to enhance MIMODet \cite{Schmid2022LowComplexity}. However, the model-driven method still remains within the scope of 
approximating Bayesian model, and its performance is somehow unpredictable when considering dynamic environment. 
By incorporating the advantage of structural input information, hybrid-learning models 
are proposed to perform online training with shallow deep neural network (DNN) 
or long short-term memory (LSTM) \cite{Cohen2022Bayesian}. The rational lies in the fact that 
structural input information will reduce the uncertainty from the environment and thus shallow NN 
is capable of capturing the variation of communication system. However, the complexity of 
online learning strategy is prohibitive in terms of the overhead control, since we need 
to perform epoch training w.r.t demodulation reference signal (DMRS) of 
each instant OFDM subframe\cite{Raviv2023Modular}. 

In this paper, we propose to connect CE and MIMODet via re-configurable AI modules where 
super resolution convolutional neural network (SRCNN) for CE training \cite{Mehran2019Deep} 
is embedded into graphical neural network (GNN) based detection process. Different from 
expectation propagation based GNN (GEPNet) method proposed in \cite{Kosasih2022Graph} and their continued  
works\cite{Zhou2023Graph2023}, where EP receiver is enhanced via GNN's output. 
We reversely enhance soft decisions by extending our previous method to provide more reliable 
prior information as input for GNN. Furthermore, the proposed method attempts to show the 
performance boundary of both reconfigurable AI based CE and MIMODet, and outperforms the 
most advanced AI based MIMODet methods in literature that we are aware of. This is verified  
via our simulation results, and also due to the fact that the proposed method performs close to 
ML detector in most scenarios. 

\section{Problem Formulation}  
\begin{figure}[!htbp]
    \centering
    \includegraphics[width = 0.48 \textwidth, keepaspectratio]{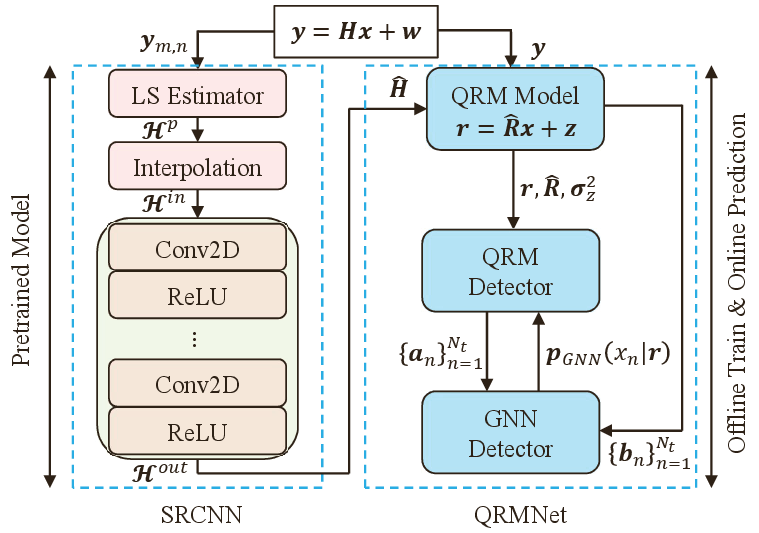}
    \caption{Paradigm of reconfigurable AI modules-based CE\&MIMODet.}  
    \label{fig:AI_CE&MIMODet}
\end{figure}

We consider a MIMO channel of size $N_{r} \times N_{t}$, where $N_{r}$ and $N_{t}$ denote the  
numbers of received and transmitted antennas, respectively. In particular, we set $N_{r}=N_{t}$ 
by default. The payloads plus the predefined demodulation reference signals (DMRS) are formatted 
into OFDM symbols via inverse fast Fourier transform (IFFT). In this way, each transmission time interval 
(TTI) contains $N_{s}$ OFDM symbols and each of them consists of $N_{c}$ subcarriers, i.e., there are 
$N=N_{c}N_{s}$ resource elements (REs) for one TTI, and $N_{p}$ of them are allocated with DMRS, and $N_{d}$ 
of them are filled with payloads.  After cyclic prefix removal and FFT operation 
upon each OFDM symbol of TTI, input and output (I/O) system model in frequency domain will be formulated 
for the purpose of CE and MIMODet. In particular, the paradigm of proposed reconfigurable AI modules for 
CE and MIMODet is depicted in Fig.~\ref{fig:AI_CE&MIMODet} to make readers grasp the technical routine 
of this paper. More details will be introduced in the following descriptions.

\subsection{SRCNN-Based Channel Estimation} 
After FFT, the received signal vector $\bm{y}_{m} \in \mathcal{C}^{N_{p} \times 1}$ w.r.t 
non-overlapped DMRS of $m$-th received antenna is expressed as 
\begin{equation}\label{eqn:system_model_ce}
    \bm{y}_{m,n} = \bm{P}_{n}\bm{h}_{m,n} + \bm{w}_{m,n}, 
\end{equation}  
where $\bm{P}_{n} \in \mathcal{C}^{N_{p} \times N_{p}}$ is a non-overlapped diagonal observation matrix 
with the $p$-th diagonal element $\{\bm{P}_{n}^{(p,p)}\}_{p=1}^{N_{p}}$ carrying the $p$-th DMRS 
of $n$-th antenna. As the target, $\bm{h}_{m,n} \in \mathcal{C}^{N_{p} \times 1}$ will be estimated 
independently per $(m,n)$-th channel link pair, and $\bm{w}_{m,n} \in \mathcal{C}^{N_{p} \times 1}$ 
denotes the noise vector that follows Gaussian independent and identical distribution (i.i.d) with a zero mean 
and and an identical variance $\sigma_{w}^{2}$.

The SRCNN based CE starts by generating the least square (LS) estimation of $\bm{h}_{m,n}$ as 
input features of neural layers by  
\begin{equation}\label{eqn:h_ls}
    \hat{\bm{h}}_{m,n}^{ls} = \left(\bm{P}_{n}^{H}\bm{P}_{n}\right)^{-1}
    \bm{P}_{n}^{H}\bm{y}_{m,n}, 
\end{equation}   
which is regarded as pre-processing stage, and the tensor $\bm{\mathcal{H}}^{p} \in \mathcal{C}^{N_{p} \times N_{r} 
\times N_{t}}$ w.r.t pilot position is constructed as 
\begin{equation}\label{eqn:H_p}
    \bm{\mathcal{H}}^{p}{\left[:,m,n\right]} = \hat{\bm{h}}_{m,n}^{ls}. 
\end{equation}

To design the architecture of SRCNN, $\bm{\mathcal{H}}^{p}$ is further interpolated 
by a Gaussian tensor $\bm{\Pi} \in \mathcal{C}^{N_{d} \times N_{p}\times N_{r} \times N_{t}}$ 
to form the tensor $\bm{\mathcal{H}}^{in} \in \mathcal{C}^{N_{d} \times N_{r} \times N_{t}}$ 
as input of SRCNN by 
\begin{equation}\label{eqn:H_in}
    \bm{\mathcal{H}}^{in} = \bm{\Pi} \odot \bm{\mathcal{H}}^{p},  
\end{equation} 
where $\odot$ denotes the operation of matrix-wise interpolation and dimension reduction on the first 
and second axis of $\bm{\Pi}$.  
The output $\bm{\mathcal{H}}^{out} \in \mathcal{C}^{N_{d} \times N_{r} \times N_{t}}$ of SRCNN  
is expressed as 
\begin{equation}\label{eqn:H_out}
    \bm{\mathcal{H}}^{out} = \mathcal{F}_{{\mathrm{SRCNN}}}\left(\bm{\mathcal{H}}^{in}; \bm{\Theta}_{\mathrm{CE}}\right), 
\end{equation} 
where $\mathcal{F}_{\mathrm{SRCNN}}\left(\cdot\right)$ represents forward propagation of SRCNN, and 
$\bm{\Theta}_{\mathrm{CE}}$ denotes the parameters of neurons to be updated during backpropagation.  
For brevity, we specified the details of $\mathcal{F}_{\mathrm{SRCNN}}\left(\cdot\right)$ in numerical simulation. 
Since we have the label tensor $\bm{\mathcal{H}}^{label} \in \mathcal{C}^{N_{d} \times N_{r} \times N_{t}}$, 
the tensor pair $(\bm{\mathcal{H}}^{out},\bm{\mathcal{H}}^{label})$ is utilized to implement 
loss function design which is a simple task and details will be omitted here. Compared to other DL based 
CE such as method introduced in \cite{Mehran2019Deep}, the proposed structure of $\mathcal{F}_{\mathrm{SRCNN}}\left(\cdot\right)$ matches the MIMODet model perfectly, i.e., one-shot 
prediction of $\bm{\mathcal{H}}^{out}$ will be applied to MIMODet w.r.t subcarriers that carry data 
for each epoch processing. It should be noticed 
that the communication system model for CE and MIMODet are in the form of complex domain, and their 
conversion to real domain for neural processing will not be elaborated either.

\subsection{QRMNet-Based Deep MIMODet}
The MIMODet assembles I/O relationship per payload subcarrier by  
\begin{equation}\label{eqn:detection_model}
    \bm{y} = \bm{H}\bm{x} + \bm{w}, 
\end{equation} 
where $\bm{y} \in \mathcal{C}^{N_{r} \times 1}$ represents the received signal collected 
over all received antennas, $\bm{H} \in \mathcal{C}^{N_{r} \times N_{t}}$ denotes the corresponding
MIMO channel per data subcarrier, $\bm{x} \in \mathcal{C}^{N_{t} \times 1}$ 
is the transmitted symbols which are drawn uniformly from a $M$-QAM constellation set 
$\mathcal{A} \triangleq [\alpha_{1}, \ldots, \alpha_{M}]^{T}$. 
MMSE criterion based QR decomposition (MMSE-QRD) \cite{Wubben2004MMSE} is leveraged 
to implement tree-structure searching by decomposing the channel matrix as 
\begin{equation}
    \tilde{\bm{H}} = \bm{Q}\bm{R}, 
\end{equation}
where $\tilde{\bm{H}} = \left[\bm{H}^{T}, \sigma\bm{I}_{N_{t}}\right]^{T} 
\in \mathcal{C}^{(N_{r}+N_{t}) \times N_{t}}$ is defined as the augmented matrix with 
$\sigma \triangleq \sigma_{w}/\sigma_{x}$, 
$\bm{Q} = \left[\bm{Q}_{1}^{T}, \bm{Q}_{2}^{T}\right]^{T}\in \mathcal{C}^{(N_{r}+N_{t}) \times N_{t}}$ 
is an orthogonal matrix, with $\bm{Q}_{1} \in \mathcal{C}^{N_{r} \times N_{t}}$ being the upper 
$(N_{r} \times N_{t})$ part of $\bm{Q}$ and $\bm{Q}_{2} \in \mathcal{C}^{N_{t} \times N_{t}}$ 
being lower $(N_{t} \times N_{t})$ part of $\bm{Q}$, and $\bm{R} 
\in \mathcal{C}^{N_{t} \times N_{t}}$ is an upper triangular matrix. By utilizing the properties 
\begin{subequations}
    \begin{align}
         \bm{I}_{N_{t}} &= \bm{Q}_{1}^{H}\bm{Q}_{1} + \bm{Q}_{2}^{H}\bm{Q}_{2}  \\  
         \bm{H} &= \bm{Q}_{1}\bm{R}    \\ 
         \bm{Q}_{2} &= \sigma\bm{I}_{N_{t}}\bm{R}^{-1}, 
    \end{align}
\end{subequations}
system model defined in \eqref{eqn:detection_model} will be rewritten as   
\begin{equation}\label{eqn:system_model_qrm}
    \bm{r} = \bm{R}\bm{x} + \bm{z}, 
\end{equation}
where we define $\bm{r} \triangleq \bm{Q}_{1}^{H}\bm{y} \in \mathcal{C}^{N_{t} \times 1}$, and $\bm{z} \triangleq \left(\bm{Q}_{1}^{H}\bm{w} 
- \sigma\bm{Q}_{2}^{H}\bm{x}\right)\in \mathcal{C}^{N_{t} \times 1}$ with the variance $\sigma_{z}^{2}$. 
The factor graph (FG)
representation of \eqref{eqn:system_model_qrm} is derived to compute the posterior 
probability of $p(\bm{x|\bm{r}})$ by approximating 
\begin{equation}\label{eqn:factor_graph_app}
    \begin{split}
        p(\bm{x}|\bm{r}) &\propto p(\bm{r}|\bm{x})p(\bm{x})     \\ 
        & \propto \exp\left(-\frac{1}{\sigma_{z}^{2}}\left\|\bm{r}-\bm{R}\bm{x}\right\|^{2}\right)p(\bm{x})  \\ 
        & \propto \exp\left(\frac{1}{\sigma_{z}^{2}}\left(2\mathrm{Re}\left\{\bm{x}^{H}\bm{c} \right\}-\bm{x}^{H}\bm{G}\bm{x}\right)\right)p(\bm{x}), 
    \end{split}
\end{equation} 
where $\bm{c} \triangleq \bm{R}^{H}\bm{r}$ and 
$\bm{G} \triangleq \bm{R}^{H}\bm{R}$. The key features of well-known 
\emph{Ungerboeck} observation model will be expressed as 
\begin{subequations}
    \begin{align}
        \bm{x}^{H}\bm{c} &= \sum\limits_{n=1}^{N_{t}}c_{n}x_{n}^{\ast}  \\ 
        \bm{x}^{H}\bm{G}\bm{x} &= \sum\limits_{n=1}^{N_{t}} G_{n,n}|x_{n}|^{2}
        + \sum\limits_{n=1}^{N_{t}}\sum\limits_{k \neq n}^{N_{t}} 2\mathrm{Re}\!\left\{G_{k,n}x_{k}x_{n}^{\ast}\right\},  
    \end{align}
\end{subequations} 
where $x_{n}$, $c_{n}$, $G_{k,n}$ denote the $n$-th element of $\bm{x}$, $n$-th element of $\bm{c}$, and 
$(k,n)$-th element of $\bm{G}$, respectively. By these definitions, we can further factorize $p(\bm{x}|\bm{r})$ 
in \eqref{eqn:factor_graph_app} as 
\begin{multline}
    p(\bm{x}|\bm{r})  \\ \propto \prod_{n=1}^{N_{t}}\underbrace{\left\{p(x_{n})F_{n}(x_{n})
        \left[\prod_{k=1, k \neq n}^{N_{t}}I_{k,n}(x_{k}, x_{n})\right]\right\}}_{p_{\mathrm{FG}}(x_{n}|\bm{r})}
\end{multline} 
with the factors  
\begin{subequations}\label{eqn:ungerboeck_model}
    \begin{align}
        & F_{n}(x_{n}) = \exp\left(\frac{1}{\sigma_{z}^{2}}\mathrm{Re}\left\{c_{n}x_{n}^{\ast}
        -\frac{1}{2}G_{n,n}\left|x_{n}\right|^{2}
        \right\}\right)      \\ 
        & I_{k,n}(x_{k},x_{n}) = \exp\left(-\frac{1}{\sigma_{z}^{2}}\mathrm{Re}\left\{G_{k,n}x_{k}x_{n}^{\ast}\right\}\right).
    \end{align}
\end{subequations}  
Considering the error of CE, key features of \emph{Ungerboeck} observations 
will be redefined by replacing $\bm{H}$ with its estimate $\hat{\bm{H}}$. Thus, the 
joint a \emph{posteriori} probability (APP) is expressed as 
\begin{multline}\label{eqn:gnn_factor_graph}
    p(\bm{x}|\bm{r}, \hat{\bm{H}}) \\ \propto 
    \prod\limits_{n=1}^{N_{t}}\left\{
    p(x_{n})\hat{F}_{n}(x_{n})
    \left[\prod_{k=1, k \neq n}^{N_{t}}\hat{I}_{k,n}(x_{k},x_{n})\right] \right\}, 
\end{multline} 
based on which GNN will be readily built to mimic the message passing (MP) between  
the factor node (FN) $\hat{I}_{k,n}(x_{k},x_{n})$ and variable node (VN) $\hat{F}_{n}(x_{n})$.   

\subsubsection{FN Update}
Based on \eqref{eqn:gnn_factor_graph}, there is always FN 
$\hat{I}_{k,n}(x_{k}, x_{n})$ to connect the variable pair of $(x_{k}, x_{n})$. 
However, different from the FN in the classic belief propagation (BP) algorithms 
that carry Gaussian messages, GNN generates features by replacing Gaussian outputs 
with DNN's prediction. Since conventional BP algorithm takes 
iteration to tackle the convergence issue, GNN will also perform iteration. In the $l$-th 
GNN iteration, the message $\bm{m}_{k,n}^{(l)}$ updated at 
$\hat{I}_{k,n}(x_{k}, x_{n})$ for $k,n \in [1, N_{t}]$, will be predicted 
by DNN $\mathcal{F}_{\mathrm{FN}}\left(\cdot\right)$, 
and its prediction is expressed by 
\begin{equation}\label{eqn:VN_to_FN}
    \bm{m}_{k,n}^{(l)} = \mathcal{F}_{\mathrm{FN}}\left(\bm{c}_{k,n}^{(l)}\right), ~k \neq n
\end{equation}   
with 
\begin{equation}\label{eqn:c_k_n}
    \bm{c}_{k,n}^{(l)} = \left[\left(\bm{u}_{k}^{(l-1)}\right)^{T}, \left(\bm{u}_{n}^{(l-1)}\right)^{T}, \bm{f}_{k,n}^{T}\right]^{T}, 
\end{equation}
where $\bm{u}_{k}^{(l-1)}$, $\bm{u}_{n}^{(l-1)}$ represent the messages aggregated 
at VNs $x_{k}$ and $x_{n}$ in the $(l-1)$-th iteration, respectively. The 
auxiliary edge attribute of $\hat{I}_{k,n}(x_{k}, x_{n})$ is defined as 
\begin{equation}\label{eqn:f_k_n}
    \bm{f}_{k,n} \triangleq \left[\hat{G}_{k,n}, \sigma_{z}^{2}\right]^{T}. 
\end{equation} 
Finally, the output $\bm{m}_{k,n}^{(l)}$ are fed back to the aggregation of VNs. 
\subsubsection{VN Update} 
The aggregation implements by first summing all incoming message
$\bm{m}_{k,n}^{(l)}$ of $x_{n}$ from its connected edges, which yields 
\begin{equation}\label{eqn:m_n_tilde}
    \tilde{\bm{m}}_{n}^{(l)} 
    = \sum\limits_{k=1, k \neq n}^{N_{t}}\bm{m}_{k,n}^{(l)}.
\end{equation} 
Since $p(x_{n}|\bm{r}, \hat{\bm{H}})$ relies on $p(x_{n})$ as the prior information   
to enhance the inference process, GNN also needs extra information $\bm{a}_{n}$ as 
partial input of neurons. The aggregated message at node $x_{n}$ is finalized by 
\begin{equation}\label{eqn:aggregated_message}
    \bm{m}_{n}^{(l)} = \left[\left(\tilde{\bm{m}}_{n}^{(l)}\right)^{T}, \bm{a}_{n}^{T}\right]^{T}.   
\end{equation}    
This is used as the input to update the messages $\bm{u}_{n}^{(l)}$ by propagating 
\begin{subequations}
    \begin{align}
        & \bm{g}_{n}^{(l)} = \mathcal{\mathcal{F}}_{\mathrm{GRU}}\left(\bm{g}_{n}^{(l-1)}, \bm{m}_{n}^{(l)}\right) \label{eqn:g_n}\\ 
        & \bm{u}_{n}^{(l)} = \mathcal{F}_{\mathrm{VN2}}\left(\bm{g}_{n}^{(l)}\right) \label{eqn:u_n}, 
    \end{align} 
\end{subequations} 
where $\mathcal{F}_{\mathrm{GRU}}\left(\cdot\right)$ is a specific gated recurrent unit (GRU) network, and 
its current and previous hidden states are denoted by $\bm{g}_{n}^{(l)}$ and $\bm{g}_{n}^{(l-1)}$, respectively.  
$\mathcal{F}_{\mathrm{VN2}}\left(\cdot\right)$ represents a DNN and the corresponding output 
$\bm{u}_{n}^{(l)}$ are utilized to update FN in \eqref{eqn:c_k_n} 
for the next iteration. In particular, $\bm{u}_{n}^{(0)}$ is initialized by 
\begin{equation}\label{eqn:u_0}
    \bm{u}_{n}^{(0)} = \mathcal{F}_{\mathrm{VN1}}\left(\bm{b}_{n}\right), 
\end{equation}   
where $\mathcal{F}_{\mathrm{VN1}}\left(\cdot\right)$ shares the same size of 
$\mathcal{F}_{\mathrm{VN2}}\left(\cdot\right)$ except for the input layer is initialized by 
\begin{equation}\label{eqn:b_n}
    \bm{b}_{n} = \left[\hat{c}_{n}, \hat{{G}}_{n,n}, \sigma_{z}^{2}\right]^{T}. 
\end{equation} 

\subsubsection{Readout Module}
After $L$ rounds of iteration between FNs and VNs in $t$-th QRMNet iteration, the readout is given by 
\begin{equation}\label{eqn:p_gnn_tilde}
    \tilde{p}_{\mathrm{GNN}}^{(t)}(x_{n}|\bm{r}) = \mathcal{F}_{\mathrm{GNN}}\left(\bm{u}_{n}^{(L)}\right), 
\end{equation} 
where $\mathcal{F}_{\mathrm{GNN}}\left(\cdot\right)$ denotes a DNN, and 
$\tilde{p}_{\mathrm{GNN}}^{(t)}(x_{n}|\bm{r})$ will be further normalized via 
softmax function, which yields 
\begin{equation}\label{eqn:p_gnn}
    p_{\mathrm{GNN}}^{(t)}(x_{n}=\alpha_{i}|\bm{r}) = \frac{\exp\left(\tilde{p}_{\mathrm{GNN}}^{(t)}\left(x_{n}=\alpha_{i}|\bm{r}\right)\right)} 
    {\sum\limits_{\alpha_{i'}\in \mathcal{A}} \exp\left(\tilde{p}_{\mathrm{GNN}}^{(t)}\left(x_{n}=\alpha_{i'}|\bm{r}\right)\right)}
\end{equation} 
for $\alpha_{i} \in \mathcal{A}$. 

\subsubsection{QRMNet Detector}
QRM detector aims to maximize the marginal posterior probability $p_{\mathrm{QRM}}(x_{n}|\bm{r})$, 
which is expressed as 
\begin{multline}\label{eqn:QRM}
        p_{\mathrm{QRM}}(x_{n}|\bm{r}) \\
        \propto \sum\limits_{\bm{x}_{\setminus x_{n}} \in \mathcal{A}^{N_{t}-1}} 
    \left(\frac{\exp\left(-\frac{\|\bm{r}-\hat{\bm{R}}\bm{x}\|^{2}}{\sigma_{z}^{2}}\right)}
    {\sum\limits_{\bm{x}' \in \mathcal{A}^{N_{t}}} \exp\left(-\frac{\|\bm{r}-\hat{\bm{R}}\bm{x}'\|^{2}}{\sigma_{z}^{2}}\right)}\right)p(x_{n}), 
\end{multline}  
where $\bm{x}_{\setminus x_{n}}$ represents all the possible combinations of 
$\mathcal{A}^{N_{t}-1}$ except for the current $n$-th antenna. 
In the $t$-th QRMNet iteration, QRM provides extra information in 
\eqref{eqn:aggregated_message} for GNN by  
\begin{equation}\label{eqn:a}
    \bm{a}_{n} 
    = \left[\hat{x}_{n}^{(t)}, \hat{v}_{n}^{(t)}\right]^{T}, 
\end{equation} 
where the mean and variance is expressed as 
\begin{subequations} 
    \begin{align}
        &\hat{x}_{n}^{(t)} = \textsf{E}\left\{x_{n}|p_{\mathrm{QRM}}^{(t)}(x_{n}|\bm{r})\right\} \label{eqn:x_n}\\ 
        &\hat{v}_{n}^{(t)} = \textsf{Var}\left\{x_{n}|p_{\mathrm{QRM}}^{(t)}(x_{n}|\bm{r})\right\} \label{eqn:v_n},  
    \end{align}
\end{subequations} 
respectively. 
In the $(t+1)$-th iteration, $p_{\mathrm{QRM}}^{(t+1)}(x_{n}|\bm{r})$ is updated by replacing 
$p(x_{n})$ in \eqref{eqn:QRM} with $p_{\mathrm{GNN}}^{(t)}(x_{n}|\bm{r})$ defined in 
\eqref{eqn:p_gnn}. Thereby, iterative detector between QRM and GNN is named as QRMNet detector. 
To better understand the structures of deployed NNs in the proposed method, Table \ref{Tab:NN_Parameters} depicts the 
specific neuron parameters in $\mathrm{Pytorch}$ for $\mathcal{F}_{\mathrm{SRCNN}}$,  $\mathcal{F}_{\mathrm{VN1}}$, 
$\mathcal{F}_{\mathrm{VN2}}$, $\mathcal{F}_{\mathrm{GRU}}$, $\mathcal{F}_{\mathrm{GNN}}$, where
neuron sizes of some intermediate layers are specified as $N_{h1}$, $N_{h2}$, $N_{h3}$, $N_{u}$, $N_{\mathcal{A}}$. 
In particular, $N_{\mathcal{A}}$ denotes the size of constellation set. 

\begin{table*}[!htbp]  
\centering 
\caption{Neurons Specification of SRCNN and QRMNET.} 
\begin{tabular}{|c|c|c|c|c|c|} \hline  
     \diagbox{stage}{item}                        &
     $\mathcal{F}_{\mathrm{SRCNN}}$ & 
     $\mathcal{F}_{\mathrm{FN}}$    & 
     $\mathcal{F}_{\mathrm{VN}}$    & 
     $\mathcal{F}_{\mathrm{GNN}}$   & 
     $\mathcal{F}_{\mathrm{GRU}}$  
     \\ \hline
     I                                          &         
     $\mathrm{Conv2d}(1,N_{h1})$                 & 
     $\mathrm{Linear}(2N_{u}+2,N_{h1})$          & 
     $\mathrm{Linear}(N_{h1},N_{h2})$            & 
     $\mathrm{Linear}(N_{u},N_{h1})$             &  
     $\mathrm{GRU}(N_{u}+N_{\mathcal{A}},N_{h1})$ 
     \\ \hline   
     II                                    &
     $\mathrm{ReLU}(~)$                      & 
     $\mathrm{ReLU}(~)$                      & 
     $\mathrm{ReLU}(~)$                      & 
     $\mathrm{ReLU}(~)$                      &   
     $\times$
     \\ \hline 
     III                                           &
     $\mathrm{Dropout}(0.3)$                         & 
     $\mathrm{Linear}(N_{h1},N_{h2})$                & 
     $\mathrm{Linear}(N_{h2},N_{h3})$                &
     $\mathrm{Linear}(N_{h1},N_{h2})$                &       
     $\times$
     \\ \hline
     IV                                              &
     $\mathrm{Conv2d}(N_{h1},N_{h2})$                & 
     $\mathrm{ReLU}(~)$                              & 
     $\mathrm{ReLU}(~)$                              & 
     $\mathrm{ReLU}(~)$                              &        
     $\times$
     \\ \hline 
     V                                             &
     $\mathrm{ReLU}(~)$                              & 
     $\mathrm{Linear}(N_{h2},N_{u})$                   & 
     $\mathrm{Linear}(N_{h3},N_{u})$                   & 
     $\mathrm{Linear}(N_{h2}, N_{\mathcal{A}})$ &       
     $\times$ 
     \\ \hline  
     VI                                         &
     $\mathrm{Dropout}(0.3)$ &$\times$ & $\times$ & $\mathrm{Softmax}(~)$ & $\times$
     \\ \hline  

     VII                                         &
     $\mathrm{Conv2d}(N_{h2},1)$ & $\times$ & $\times$ & $\times$ & $\times$                         
     \\ \hline 
     
\end{tabular} 
\label{Tab:NN_Parameters} 
\end{table*}

\begin{table*}[!htbp]  
\centering 
\caption{Complexity Analysis of QRMNET w.r.t Multiplications.} 
\begin{tabular}{|c|c|c|c|c|c|} \hline  
     \diagbox{stage}{item}                        &
     $\mathcal{F}_{\mathrm{SRCNN}}$ & 
     $\mathcal{F}_{\mathrm{FN}}$    & 
     $\mathcal{F}_{\mathrm{VN}}$    & 
     $\mathcal{F}_{\mathrm{GNN}}$   & 
     $\mathcal{F}_{\mathrm{GRU}}$  
     \\ \hline
     I                                          &         
     $81N_{h1}(N_{s}-8)(N_{c}-4)$                 & 
     $2(N_{u}+1)(N_{t}-1)N_{h1}N_{t}$          & 
     $N_{t}N_{h1}N_{h2}$            & 
     $N_{t}N_{u}N_{h1}$             &  
     $N_{t}(N_{u}+N_{\mathcal{A}}+N_{h1})N_{h1}$ 
     \\ \hline   
     II                                    &
     $\times$                      & 
     $\times$                      & 
     $\times$                      & 
     $\times$                      &   
     $\times$
     \\ \hline 
     III                                           &
     $\times$                         & 
     $N_{t}(N_{t}-1)N_{h1}N_{h2}$                & 
     $N_{t}N_{h2}N_{h3}$                &
     $N_{t}N_{h1}N_{h2}$                &       
     $\times$
     \\ \hline
     IV                                              &
     $N_{h1}N_{h2}N_{c}(N_{s}-4)$                                  & 
     $\times$                              & 
     $\times$                              & 
     $\times$                              &        
     $\times$
     \\ \hline 
     V                                             &
     $\times$                              & 
     $N_{t}(N_{t}-1)N_{h2}N_{u}$                   & 
     $N_{t}N_{h3}N_{u}$                   & 
     $N_{t}N_{h2}N_{\mathcal{A}}$ &       
     $\times$ 
     \\ \hline  
     VI                                         &
     $\times$ &$\times$ & $\times$ & $\times$ & $\times$
     \\ \hline  

     VII                                         &
     $N_{h2}N_{c}(N_{s}-4)$ & $\times$ & $\times$ & $\times$ & $\times$                         
     \\ \hline  
     
\end{tabular} 
\label{Tab:Complexity} 
\end{table*}

\subsection{Algorithm Summary}  
In addition to Fig.~\ref{fig:AI_CE&MIMODet} which has illustrated how SRCNN and QRMNet coexist 
to achieve the AI based CE and MIMODet, a more detailed implement procedure is provided in Alg.~\ref{Alg:SRCNN_QRMNet} 
to demonstrate the training phases of SRCNN and QRMNet, respectively. The iteration number of
QRMNet can be set as $t=1$ to reduce the algorithm's complexity.  
\begin{algorithm}
    \begin{algorithmic}[1]
        \caption{Reconfigurable AI modules for CE\&MIMODet} 
        \label{Alg:SRCNN_QRMNet} 
        \STATE {\bf Pre-trained SRCNN-CE} 
        \STATE {\bf Input:} received signal $\bm{y}_{m,n}$, interpolation tensor $\bm{\Pi}$. 
        \STATE compute $\hat{\bm{h}}_{m,n}^{ls}$ using \eqref{eqn:h_ls}.  
        \STATE compute $\bm{\mathcal{H}}^{p}$ and $\bm{\mathcal{H}}^{in}$ using \eqref{eqn:H_p}  
               and \eqref{eqn:H_in}, respectively.  
        \STATE offline train $\mathcal{F}_{\mathrm{SRCNN}}(\cdot)$ using \eqref{eqn:H_out}. 
        \STATE predict $\bm{\mathcal{H}}^{out}$ and generate $\{\hat{\bm{H}}\}_{d=1}^{N_{d}}$ used for MIMODet.  
        \STATE {\bf Offline Procedure of QRMNet-MIMODet}
        \STATE {\bf Input:} received signal $\bm{y}$ and predicted channel $\hat{\bm{H}}$.
        \STATE compute $\bm{r}$, $\hat{\bm{R}}$, $\sigma_{z}^{2}$ for QRM detector detector   
        \STATE initialize $\{\bm{a}_{n}\}_{n=1}^{N_{t}}$ by QRM using \eqref{eqn:a}. 
        \STATE compute $\{\hat{c}_{n}\}_{n=1}^{N_{t}}, \{\hat{G}_{n,n}\}_{n=1}^{N_{t}}$ to construct $\{\bm{b}_{n}\}_{n=1}^{N_{t}}$ using \eqref{eqn:b_n}.   
        \STATE compute $\{\bm{f}_{k,n}\}$, $\forall k, n \in [1:N_{t}]$, $k \neq n$ using \eqref{eqn:f_k_n}
        \FOR {$t=1, \ldots, T$} 
        \STATE implement MLP to compute $\{\bm{u}_{n}^{(0)}\}_{n=1}^{N_{t}}$ using \eqref{eqn:u_0}.  
        \FOR {$l=1,\ldots,L$}   
        \STATE {\bf FN update}:    
        \STATE compute $\bm{c}_{k,n}^{(l)}$, $\forall k, n \in [1:N_{t}]$, $k \neq n$ using \eqref{eqn:c_k_n}. 
        \STATE implement MLP to compute $\bm{m}_{k,n}^{(l)}$, $\forall k, n \in [1:N_{t}]$, $k \neq n$ 
        using \eqref{eqn:VN_to_FN}. 
        \STATE {\bf VN update}: 
        \STATE compute $\{\tilde{\bm{m}}_{n}^{(l)}\}_{n=1}^{N_{t}}$ using \eqref{eqn:m_n_tilde}   
        \STATE compute $\{\bm{m}_{n}^{(l)}\}_{n=1}^{N_{t}}$ using \eqref{eqn:aggregated_message}   
        \STATE implement MLP to compute $\{\bm{g}_{n}^{(l)}\}_{n=1}^{N_{t}}$ and 
               $\{\bm{u}_{n}^{(l)}\}_{n=1}^{N_{t}}$ 
               using \eqref{eqn:g_n} and \eqref{eqn:u_n}, respectively.  
        \ENDFOR 
        \STATE {\bf Readout}:  
        \STATE implement MLP to compute $\tilde{p}_{\mathrm{GNN}}^{(t)}(x_{n}|\bm{r})$ using \eqref{eqn:p_gnn_tilde} 
        \STATE compute $\{p_{\mathrm{GNN}}^{(t)}(x_{n}=\alpha_{i}|\bm{r})\}_{i=1}^{M}$ using \eqref{eqn:p_gnn}.    
        \STATE {\bf QRM enhancement}:
        \STATE enhance QRM detector using $p_{\mathrm{GNN}}^{(t)}(x_{n}|\bm{r})$. 
        \STATE compute $\{\hat{x}_{n}^{(t)}\}_{n=1}^{N_{t}}$ and $\{\hat{v}_{n}^{(t)}\}_{n=1}^{N_{t}}$ using 
               \eqref{eqn:x_n} and \eqref{eqn:v_n}, respectively. 
        \ENDFOR  
        \STATE preserve $\mathcal{F}_{\mathrm{SRCNN}}$, $\mathcal{F}_{\mathrm{VN1}}$, $\mathcal{F}_{\mathrm{VN2}}$, 
               $\mathcal{F}_{\mathrm{GRU}}$, $\mathcal{F}_{\mathrm{GNN}}$ for online testing procedure of QRMNet-MIMODet.  
        \STATE {\bf Online Procedure of QRMNet-MIMODet.}  
        \STATE the reserved AI neurons are reused to predict new payloads.  
    \end{algorithmic} 
\end{algorithm}
We evaluate the number of multiplication operation for online 
phase of QRMNet. This evaluation guides us select appropriate neuron size to achieve the tradeoff between  
performance and complexity. Table ~\ref{Tab:Complexity} provides our calculation of multiplication 
operation in each stage with 
regards to neuron parameters in Tab.~\ref{Tab:NN_Parameters}. In particular,  
$\mathrm{kernel\_size}$ of $\mathrm{Conv2d}$ in 
$\mathcal{F}_{\mathrm{SRCNN}}(\cdot)$ are $9,1,5$, respectively, and padding size is set by 2 across all 
convolutional layers. Activation functions such as 
$\mathrm{ReLU}$ and $\mathrm{Softmax}$ are not considered in the analysis. The total complexity of 
neurons' multiplication for is approximated to $\mathcal{O}(N_{1}N_{h2}N_{c}N_{s}
+LN_{t}^{2}(N_{u}N_{h1}+N_{u}N_{h2}+N_{h1}N_{h2})+N_{t}N_{h1}(N_{u}+N_{\mathcal{A}}+N_{h1}))$ for each outer iteration.  
\section{Numerical Simulation}  
The simulation evaluates the performance of our proposed method. In particular, $\mathrm{Pytorch}$ AI package 
with NVIDIA GeForce RTX4090 GPU@32GB platform is used for AI implementation. 
Table \ref{Tab:Recong_AI_Param} shows the system-level configuration for simulation. 
The propagation channel models are generated according to 3GPP standard \cite{3GPPChannelModel} 
with Rayleigh distribution. The generated multipath channels are converted to frequency domain 
$\bm{H}$ per subcarrier with different spatial correlation levels which are characterized by 
$\alpha$ and $\beta$. LDPC encoder/decoder with 1/2- 
and 1/3 coding rate is evaluated. We compare BLER of proposed QRMNet with baselines such as 
expectation propagation (EP) \cite{Kosasih2022Graph} and QRM detectors \cite{Wubben2004MMSE}, 
where LMMSE with perfect measurements of mean vector and covariance matrix is utilized for CE. 
As the counterpart of reconfigurable reconfigurable modules itself, QRM's replacement by EP 
to provide priors to enhance GNN is added as the AI baseline to demonstrate the superiority  
of QRMNet to GEPNet, which is the best state of the art of AI based MIMODet\cite{Kosasih2022GraphJ}. 
Therefore, QRMNet needn't to be compared with other AI baselines. 
\begin{table}[!htbp] 
    \centering
    \caption{Parameters of system configuration.} 
    \begin{tabular}{|c|c|c|}                                                                      \hline 
         Parameters &  Description & Value                                                               \\  \hline 
         $f_{s}$               & sampling frequency                      & $960$ (kHz)            \\  \hline            
         $N_{f}$               & FFT size                                & $64$                   \\  \hline   
         $N_{c}$               & number of subcarriers/OFDM symbol       & $48$                   \\  \hline 
         $N_{s}$               & number of OFDM symbols/TTI              & $14$                   \\  \hline 
         $N_{p}$               & number of pilots/TTI                    & $32$                   \\  \hline 
         $N_{d}$               & number of data/TTI                      & $480$                  \\  \hline 
         $N_{b}$               & number of TTI/epoch                     & $2000$                 \\  \hline  
         $N_{e}$               & number of epochs (training)             & $300$                  \\  \hline 
         $T$                   & iter. of QRMNet or GEPNet               & $2$                    \\  \hline 
         $L$                   & iter. of GNN                            & $10$                   \\  \hline 
         $\alpha$              & tx correlation level                    & $0-0.3$                \\ \hline 
         $\beta$               & rx correlation level                    & $0-0.3$                \\ \hline 
         $\mathcal{A}$         & QAM size                                & 16, 64, 256            \\ \hline  
         $N_{r} \times N_{t}$  & MIMO size                               & $2 \times 2$, $4 \times 4$    \\ \hline  
         $N_{u}$               & neuron size                             & $8$                \\ \hline 
         $N_{h1}$              & neuron size                             & $64$               \\ \hline 
         $N_{h2}$              & neuron size                             & $32$               \\ \hline 
         $N_{h3}$              & neuron size                             & $64$               \\ \hline 
    \end{tabular}
    \label{Tab:Recong_AI_Param} 
\end{table}   

\begin{figure}[!htbp]
    \centering
    \includegraphics[width = 0.48 \textwidth, keepaspectratio]{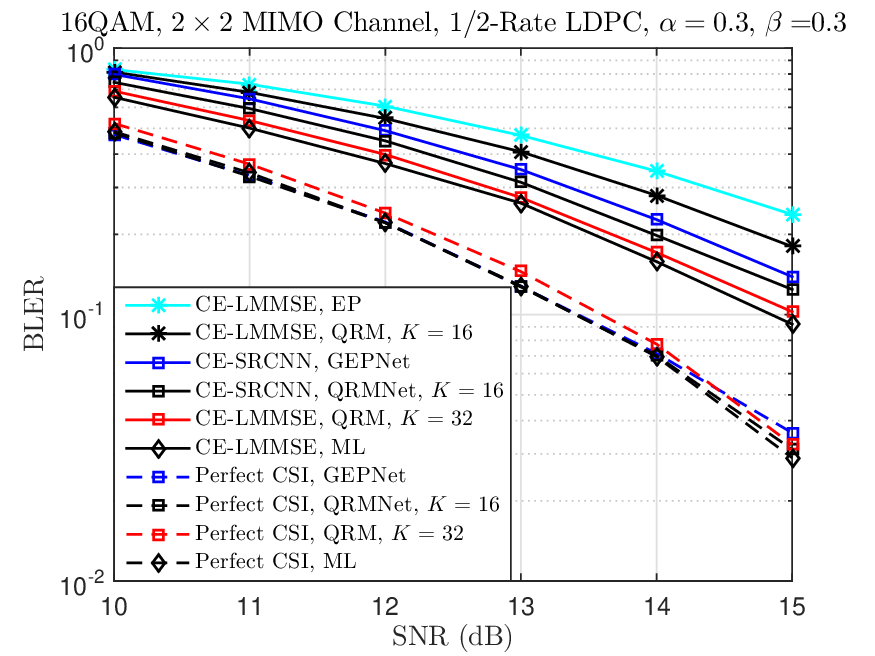}
    \caption{BLER performance with $2 \times 2$ MIMO, 16QAM.}    
    \label{fig:BLER_AI_CE_SD_16QAM_2Tx_2Rx_Alpha0.3_Beta0.3}
\end{figure}

We present the simulation results from lower modulation and medium-high level 
channel's correlation coefficients with 16QAM, $2 \times 2$ MIMO channel, $\alpha=0.3$, $\beta=0.3$. 
Figure \ref{fig:BLER_AI_CE_SD_16QAM_2Tx_2Rx_Alpha0.3_Beta0.3} shows the BLER of proposal method
compared to other baselines. Under perfect CSI assumption, QRMNet with $K=16$ survival paths  
approaches to QRM with $K=32$ survival paths and ML, which means that QRMNet can achieve the ML 
performance while preserving fewer survival paths compared to QRM detector. When considering CE 
error, SRCNN CE based QRMNet with $K=16$ performs better than LMMSE-CE based QRM with $K=16$ by 
$0.5$ dB thanks to GNN's enhancement to MIMODet. It only has $0.5$ dB performance loss compared 
to LMMSE CE based ML detector and performs better than SRCNN CE based GEPNet. It's also observed 
that SRCNN CE based QRMNet has $1.5$ dB performance gain than LMMSE CE based EP detector.
\begin{figure}[!htbp]
    \centering
    \includegraphics[width = 0.48 \textwidth, keepaspectratio]{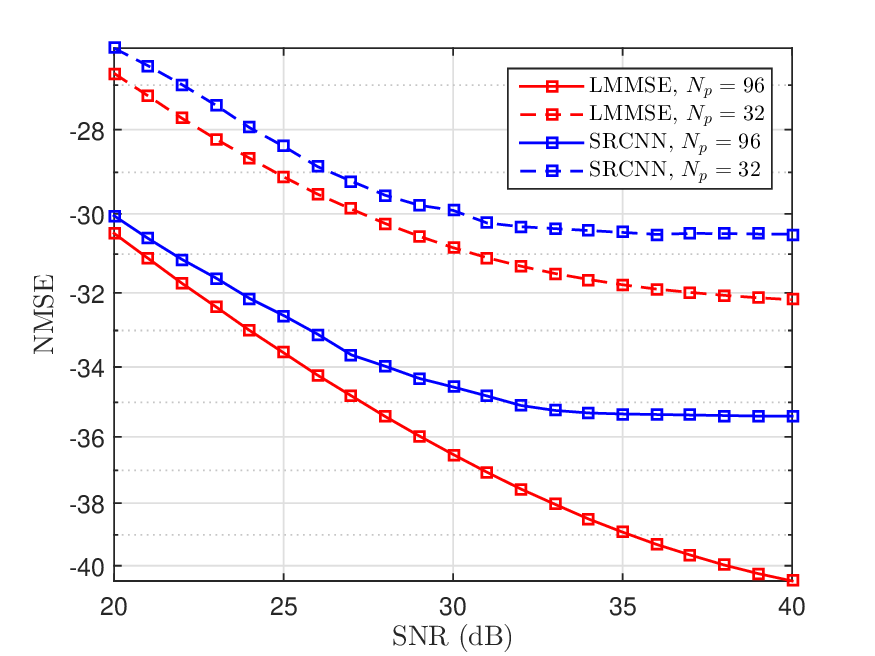}
    \caption{NMSE of CE with $\alpha=0.3$, $\beta=0.3$.}    
    \label{fig:NMSE_CE}
\end{figure}  
Furthermore, Fig.~\ref{fig:NMSE_CE} shows that NMSE of SRCNN saturates after $32$ dB and performs  
worse than LMMSE CE due to the underfitting incapability of $\mathcal{F}_{\mathrm{SRCNN}}$. It 
explains the reason why LMMSE CE based QRM with $K=32$ works slightly better than 
SRCNN CE based QRMNet with $K=16$ in Fig.~\ref{fig:BLER_AI_CE_SD_16QAM_2Tx_2Rx_Alpha0.3_Beta0.3}.  
It should be noticing that the performance of SRCNN CE can be further improved by considering a 
more complex AI structure in the future study, which will not discussed in this section. 

\begin{figure}[!htbp]
    \centering
    \includegraphics[width = 0.48 \textwidth, keepaspectratio]{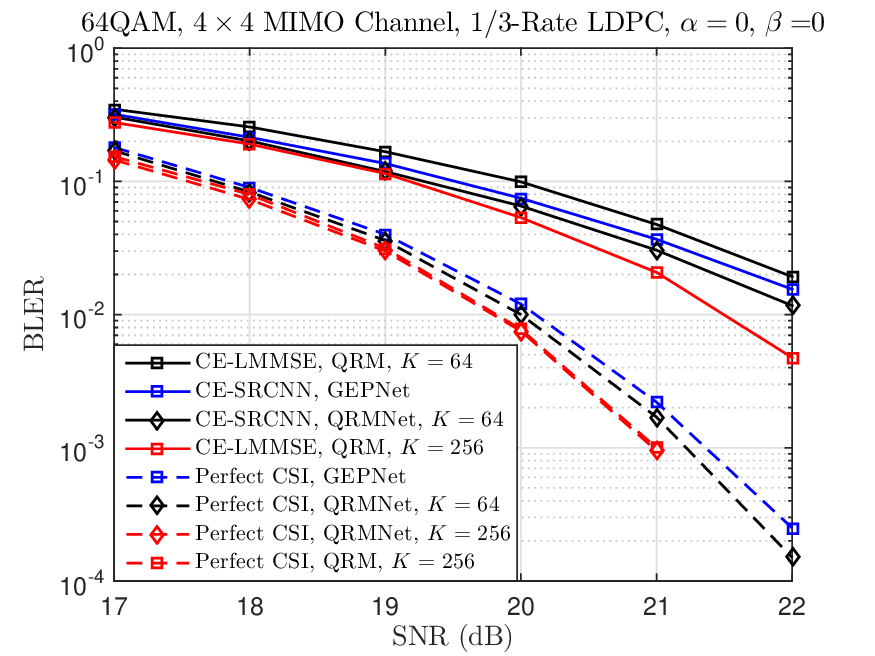}
    \caption{BLER performance with $4 \times 4$ MIMO, 64QAM.}    
    \label{fig:BLER_AI_CE_SD_64QAM_4Tx_4Rx_Alpha0_Beta0}
\end{figure}

Figure \ref{fig:BLER_AI_CE_SD_64QAM_4Tx_4Rx_Alpha0_Beta0} shows the results  
with 64QAM, $4 \times 4$ MIMO, and low-level channel's correlation. Observations 
of Fig.~\ref{fig:BLER_AI_CE_SD_16QAM_2Tx_2Rx_Alpha0.3_Beta0.3} still applies to 
Fig.~\ref{fig:BLER_AI_CE_SD_64QAM_4Tx_4Rx_Alpha0_Beta0}, where SRCNN CE based QRMNet 
with $K=64$ outperforms LMMSE CE based QRM with $K=64$ 
when considering CE error. It also has $0.5$ dB performance loss compared to 
LMMSE CE based QRM with $K=256$. Performance gain of QRMNet can be obtained by increasing 
the survival paths, e.g., QRMNet with $K=256$ has the same BLER as QRM with $K=256$ achieved. 
It means that GNN cannot further enhance the MIMO when priors from QRM 
is statistically sufficient. Nevertheless, SRCNN CE based QRMNet still performs better 
than SRCNN CE based GEPNet which is considered as the best AI receiver as concerned. 


\begin{figure}[!htbp]
    \centering
    \includegraphics[width = 0.48 \textwidth, keepaspectratio]{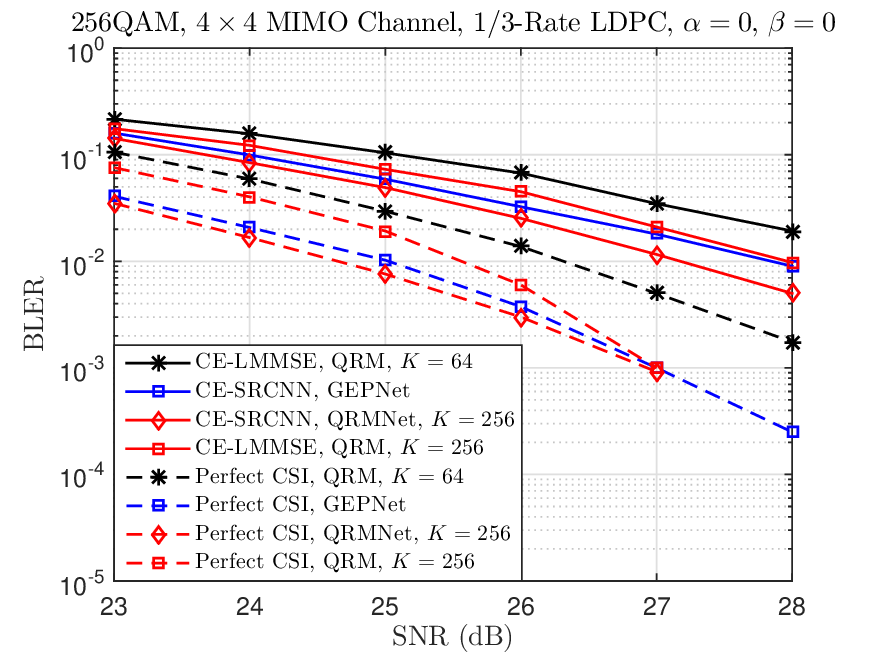}
    \caption{BLER performance with $4 \times 4$ MIMO, 256QAM.}    
    \label{fig:BLER_AI_CE_SD_256QAM_4Tx_4Rx_Alpha0_Beta0}
\end{figure}

Figure \ref{fig:BLER_AI_CE_SD_256QAM_4Tx_4Rx_Alpha0_Beta0} shows results 
when baseline QRM takes $K\leq256$. Both SRCNN CE based- 
QRMNet and GEPNet outperforms LMMSE CE based QRM baseline with $K=256$, which implies that GNN's 
enhancement is more significant when referring higher modulation order and limited 
number of survival paths for QRM baseline. 

\section{Conclusions}  
We have proposed reconfigurable AI aided method for CE and MIMODet. In general, the proposal performs 
close to the baseline of LMMSE-interpolated CE combined with QRM MIMODet under low and medium 
correlated channels, and is better than the QRM baseline when the survival paths are insufficient 
to achieve optimal detection. Although SRCNN based CE yields a higher error-floor than 
LMMSE-interpolation CE, a more sophisticated AI module can easily enhance it. Moreover, simulation 
results have shown that GNN as the AI component for MIMODet, can be enhanced via statistical priors 
from conventional QRM or EP algorithms. Future research directions include to fully exploit the 3D based CE, 
and also jointly train the CE and MIMO detector aiming for a better data-detection performance.

\bibliographystyle{IEEEtran}

\end{document}